\documentclass{aastex631}

\usepackage{lipsum}
\usepackage{amsmath}
\usepackage{booktabs}

\newcommand{\dd}{\mathrm{d}}
\newcommand{\ddx}[2]{\frac{\mathrm{d}#1}{\mathrm{d}#2}}

\newcommand{\unit}[1]{\,\mathrm{#1}}
\newcommand{\hh}[0]{\mathrm{H_2}}
\newcommand{\nn}[0]{\mathrm{N_2}}

\newcommand{\wa}[0]{\mathrm{H_2O}}
\newcommand{\cd}[0]{\mathrm{CO_2}}
\newcommand*{\rom}[1]{\uppercase\expandafter{\romannumeral #1\relax}}

\begin{document}
\title{Bistability, Oscillations, and Multistability on Hycean Planets}
\author{Yichen Gao}
\affiliation{Department of Atmospheric and Oceanic Sciences, School of Physics, Peking University}

\author{Daniel D. B. Koll}
\affiliation{Department of Atmospheric and Oceanic Sciences, School of Physics, Peking University}

\author{Feng Ding}
\affiliation{Department of Atmospheric and Oceanic Sciences, School of Physics, Peking University}

\begin{abstract}
Hycean planets are hypothetical exoplanets characterized by $\wa$ oceans and $\hh$-rich atmospheres. These planets are high-priority targets for biosignature searches, as they combine abundant surface liquid water with easy-to-characterize $\hh$-rich atmospheres. Perhaps their most unusual climate feature is convective inhibition, which can dramatically alter a planet's temperature structure. However, so far hycean planets have mostly been investigated using 1D models that do not account for convective inhibition, and its effects are still poorly understood. This work develops pen-and-paper theory to analyze the effects of moist convective inhibition on hycean planets. The theory is tested and verified against a 1D radiative-convective model. We show that hycean planets near the onset of convective inhibition can exhibit either bistability or oscillations, due to the inhibition layer's trapping of heat and moisture. Meanwhile, hot hycean planets exhibit multistability, in which the inhibition layer and surface climate show multiple stable equilibria due to the lack of constraints on the water cycle inside the inhibition layer. The water cycle inside the inhibition layer is influenced by numerous processes that are challenging to resolve in 1D, including turbulent diffusion, convective overshoot and large-scale circulations. Our results demonstrate that hycean planets have unexpectedly rich climate dynamics. Meanwhile, previous claims about hycean planets should be treated with caution until confirmed with more self-consistent 1D and 3D models; this includes the claim that K2-18b might be habitable, and the proposal to infer $\wa$ oceans on sub-Neptunes from JWST measurements of chemical species in their upper atmospheres.
\end{abstract}


\section{Introduction}

Hycean planets are a class of hypothetical exoplanets characterized by water oceans and hydrogen-rich atmospheres. Candidates for such planets are low-density water-rich sub-Neptunes  \citep{madhusudhan_habitability_2021}, such as K2-18b \citep{madhusudhan_carbon-bearing_2023}, LHS-1140b \citep{cadieux_new_2024, damiano_lhs_2024}, and TOI-270d \citep{holmberg_possible_2024}. Their atmospheres have distinctively large scale heights due to $\hh$'s low molecular weight. This results in a larger signal-to-noise ratio in transmission spectra, making biosignatures, if any, easier to detect \citep{madhusudhan_habitability_2021}. However, there is still significant debate about the internal structures of these low-density water-rich sub-Neptunes \citep{wogan_jwst_2024, cadieux_transmission_2024}, particularly whether these planets' high-pressure atmospheres still allow the underlying water ocean to remain liquid or whether any water layer would be super-critical. 

An alternative scenario for hycean planets is water-rich super-Earths (including water worlds) with $\hh$-rich atmospheres.
An $\hh$ atmosphere can enable such planets to maintain a liquid ocean in a more distant orbit than planets with an $\nn$-$\cd$ atmosphere, which means the habitable zone of hycean planets extends much farther out from their host star \citep{pierrehumbert_hydrogen_2011}. By virtue of having thinner atmospheres than more gas-rich sub-Neptunes, these super-Earths are more likely to have sub-critical oceans with liquid surfaces, but face the problem of losing their hydrogen atmospheres over time. Super-Earths with $\hh$ atmospheres are thus particularly attractive targets at larger planet-star separations, where atmospheric escape is less effective, making them good targets for future direct imaging missions \citep{pierrehumbert_hydrogen_2011}. 
Altogether, both temperate sub-Neptunes and long-period super-Earths could potentially support hycean climates, making them attractive targets for biosignature searches.

One important way in which hycean planets with $\hh$-$\wa$ atmospheres differ from traditionally habitable planets with $\nn$-$\cd$-$\wa$ atmospheres is their atmospheric convection. Due to the low molecular weight of $\hh$, water vapor can become enriched in the lower atmosphere, creating a mean molecular weight gradient that suppresses convection. The criterion for convective inhibition is the Ledoux criterion $\nabla_T<\nabla_{ad}+\nabla_\mu$, where $\nabla_T=\ddx{\ln T}{\ln p}$ is the thermal gradient, $\nabla_{ad}=\ddx{\ln T}{\ln p}|_{ad}$ is the adiabatic temperature gradient, and $\nabla_\mu=\ddx{\ln T}{\ln \mu}$ is the mean molecular weight gradient \citep{ledoux_stellar_1947}. A higher molecular weight gradient requires a higher thermal gradient to trigger convection. In the case of moist convection, the mean molecular weight is directly linked to temperature by the Clausius-Clapeyron relation, and the Ledoux criterion becomes
\begin{equation}
    (\nabla_T-\nabla_{ad}^*)(q_v-q_{cri})<0,
\end{equation}
where $\nabla_{ad}^*$ is the moist adiabatic temperature gradient, $q_v$ is the vapor concentration, and
\begin{equation}\label{eq:q_cri}
    q_{cri} = \frac{R^*T}{(M_v-M_d)L}.
\end{equation}
When $q_v>q_{cri}$, moist convection is inhibited in a superadiabatic atmosphere \citep{guillot_condensation_1995, leconte_condensation-inhibited_2017}, leading to a convection regime completely different from that on Earth. The threshold $q_v>q_{cri}$ has also been called the ``Guillot threshold'' \citep{seeley_resolved_2025}. In Equation \ref{eq:q_cri}, $R^*$ is the universal gas constant, $T$ is temperature, $L$ is the specific latent heat of the phase transition, and $M_v$ and $M_d$ are the molecular weights of water vapor and dry air.

The concept of moist convective inhibition has been widely explored for giant planets in the solar system. The observed abundances of $\wa$ in Jupiter and Saturn and $\mathrm{CH_4}$ in Uranus and Neptune are close to $q_{cri}$ from Equation \ref{eq:q_cri} \citep{leconte_condensation-inhibited_2017}. The resulting moist convective inhibition can have large consequences; for example, Saturn's great white spot has been explained as a recurring convective storm driven by moist convective inhibition \citep{li_moist_2015}, with similar storms likely on Uranus and Neptune \citep{clement_storms_2024}. By modulating internal heat fluxes, convective inhibition also strongly influences the giant planets' long-term thermal evolution \citep{friedson_inhibition_2017, markham_constraining_2021}. 
Recent work has thus been developing better theoretical and numerical models to explore the impact of moist convective inhibition. One of the most distinctive features is that a convectively inhibited atmosphere develops a distinct multi-layer structure, with a radiative inhibition layer around the cloud level which separates the upper (moist-adiabatic) and lower (dry-adiabatic) parts of the atmosphere \citep{leconte_condensation-inhibited_2017,leconte_3d_2024}.

On hycean planets convective inhibition should be similarly important, but most work in this area has relied on 1D models that ignore inhibition effects \citep[e.g.,][]{piette_temperature_2020,madhusudhan_interior_2020,madhusudhan_habitability_2021,nicholls_agni_2025}. For example, multiple theoretical papers have argued that telescope observations of sub-Neptunes can tell whether these planets have liquid water oceans \citep{yu_how_2021,hu_unveiling_2021}. The idea is that the abundances of chemical species such as CO$_2$, CH$_4$ and NH$_3$ in the upper atmosphere of sub-Neptunes should strongly differ between planets with liquid water surfaces versus planets with thick gas envelopes. However, none of these proposals considered the potential effects of moist convective inhibition. In reality, once an atmosphere becomes inhibited, its atmosphere stops being convective and well-mixed so the chemical state of the upper atmosphere might no longer closely reflect the planet's interior.
Similarly, it is still unclear how to interpret the \textit{James Webb Space Telescope} (JWST) observations of sub-Neptunes like K2-18b and TOI-270d. One interpretation holds that these planets are hycean planets, featuring relatively thin $\mathcal{O}(1-100)$ bar $\hh$-rich atmospheres above liquid $\wa$ oceans \citep{madhusudhan_habitability_2021,holmberg_possible_2024,hu_water-rich_2025}, another is that these planets have thick gaseous atmospheres, potentially coupled to a deep magma ocean \citep{wogan_jwst_2024,shorttle_distinguishing_2024,hu_water-rich_2025}. However, both interpretations are based on models that do not account for the effects of convective inhibition.

How does convective inhibition affect hycean planets? Unfortunately, work in this area is still relatively limited. \cite{markham_convective_2022} investigated the effect on convective inhibition on the interiors and longterm evolution of hycean planets.
Turning to surface climates, \cite{innes_runaway_2023} used a 1D radiative-convective model in which the atmosphere was assumed to be fully saturated. They showed that the onset of moist convective inhibition speeds up the development of the runaway greenhouse, significantly moving the inner edge of the hycean habitable zone further from the star compared to models that do not account for moist convective inhibition \citep{koll_hot_2019}.
Multiple papers considered small-scale convection in hycean atmospheres using 3D cloud-resolving models \citep{leconte_3d_2024,habib_3d_2024,seeley_resolved_2025}. These works confirm the basic prediction from \citet{guillot_condensation_1995}, namely that once $q_v>q_{cri}$, the atmosphere's structure changes dramatically and a super-adiabatic layer develops near the surface. However, the same papers also showed unexpected results. For example, \citet{seeley_resolved_2025} found that, in some of their simulations, convection becomes episodic, potentially mimicking Saturn's storms \citep{li_moist_2015}. Similarly, \citet{habib_3d_2024} found that convective adjustment in $\hh$-$\wa$ atmospheres depends on the initial scale of perturbations, such that small-scale perturbations lead to different outcomes than large-scale ones.
So far, these investigations have been based on numerical models, leaving many open questions: When does a hycean atmosphere transition from fully convective to inhibited? When is moist convection episodic, when steady? Why does perturbation size matter? And are there any other relevant thresholds besides the Guillot threshold (Equation \ref{eq:q_cri})?

The goal of this work is to develop pen-and-paper theory to better understand moist convection on hycean planets. The theory is verified using 1D numerical simulations. We focus on relatively thin atmospheres (surface pressure $<\mathcal{O}(10)\unit{bar}$), so the inhibition layer formed near the surface still affects the planet's outgoing longwave radiation (OLR). 
First, we re-examine the transition between convective and inhibited states based on a simple thought experiment in which a hycean atmosphere is heated up or cooled down by varying the incoming stellar flux. We show that the critical flux at which convection becomes inhibited while heating up is different from the critical flux at which convection resumes again while cooling down.
As a result, hycean planets close to the Guillot threshold either exhibit bistability (the atmosphere is in one of two stable equilibrium states, inhibited or convective) or oscillations (the atmosphere has no stable equilibrium and cycles between inhibited and convective).
Second, for hot hycean planets significantly above the Guillot threshold, we report multistability. The atmosphere's vertical structure stops being uniquely defined by the top-of-atmosphere (TOA) energy budget, and the convective inhibition layer and surface temperature can be in any one of multiple stable equilibria. Moreover, in this regime, the vertical structure becomes extremely sensitive to the diffusion of water vapor through the inhibition layer, which is poorly constrained.

The rest of this paper is as follows. In Section \ref{sec:theory}, we develop the theory for the onset versus breakdown of convective inhibition. Section \ref{1dmodel} describes the one-dimensional numerical model used to test our theory. Section \ref{sec:sim_rslt} analyzes the numerical results and confirms our theoretical predictions, while further highlighting the potential for multistability. Finally, Section \ref{sec:dis} summarizes our results and discusses their implications.

\section{Theory}\label{sec:theory}

\subsection{The Transition Between Convective and Inhibited States}\label{Sec:tran}

\begin{figure}[t]
    \centering
    \includegraphics[width=0.7\linewidth]{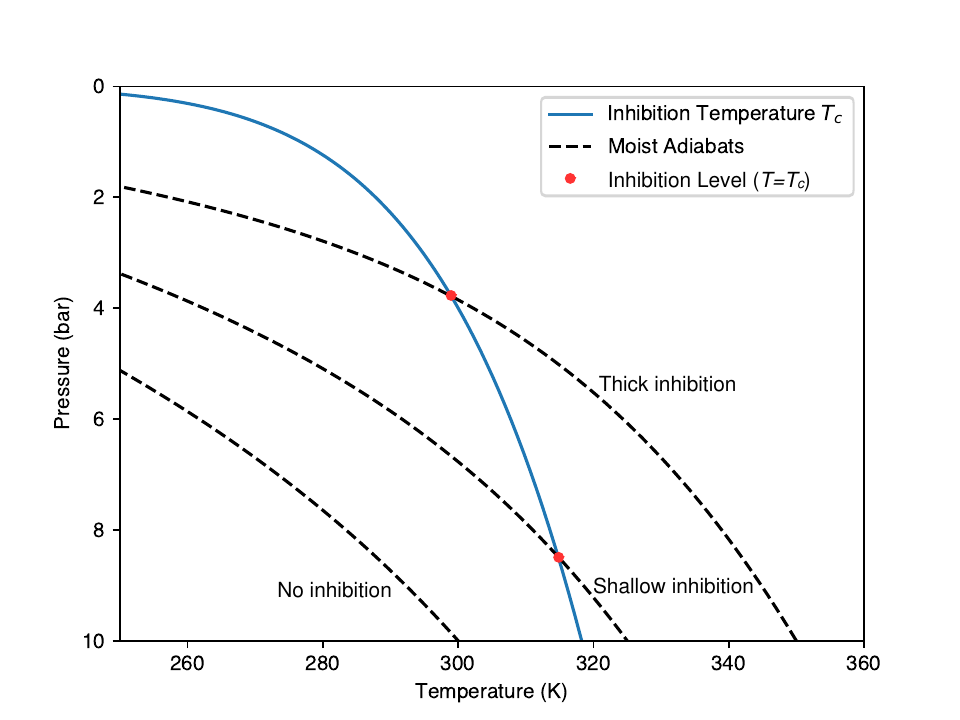}
    \caption{Hycean atmospheres become inhibited from the bottom-up; as the planet warms, the inhibition layer first forms right above the surface. Solid line shows the Guillot threshold temperature $T_c$. Dashed lines are moist adiabats with surface temperature equal to $300\unit{K}$, $325\unit{K}$, and $350\unit{K}$, from left to right, respectively. The intersection $T=T_c$ represents the approximate inhibition level, below which moist convection will be inhibited. The assumed parameters are $p_s=10\unit{bar}$, and $\wa$ convection in background air with $M_d=2.3\unit{g/mol}$.}
    \label{fig:Tc-vs-p}
\end{figure}

Consider the following thought experiment. Start with a hycean planet cool enough so that its atmosphere and surface fall below the Guillot threshold (Equation \ref{eq:q_cri}). For convenience, we rewrite the Guillot threshold as a temperature criterion; assuming that the atmosphere is saturated, the onset temperature for convective inhibition is
\begin{equation}\label{eq:Tc}
    T_c=\frac{(M_v-M_d)L(T_c)}{R^*}q_{sat}(p,\,T_c),
\end{equation}
where $q_{sat}=\frac{M_v}{M_v+M_d(p/e(T_c)-1)}$ is the specific humidity at saturation, $p$ is the pressure, and $e(T_c)$ is the saturation vapor pressure. As long as $T<T_c$ everywhere, the planet's atmosphere will thus be in standard radiative-convective equilibrium.

Next, imagine the planet first warms up and then cools down again, say, by increasing and decreasing the planet's received stellar flux. As the planet warms up, which part of the atmosphere becomes inhibited first? Figure \ref{fig:Tc-vs-p} shows the temperature-pressure profiles of three moist adiabats, with surface temperature ranging from $300\unit{K}$ to $350\unit{K}$, and $T_c(p)$ based on Equation \ref{eq:Tc}. The slope of the $T_c$ curve is always steeper than any moist adiabatic slopes. As a result, as the atmosphere warms, the inhibition layer starts to form at the bottom of the atmosphere. Below the intersection of the atmosphere's temperature-pressure profile with the $T_c$ curve, moist convection is inhibited. We call the case where such an intersection exists (surface temperature $T_s>T_c(p_s)$, where $p_s$ is the surface pressure) an inhibited state and the case where such an intersection does not exist ($T_s<T_c(p_s)$) a fully-convective, or just convective, state.

\begin{figure}[t]
    \centering
    \includegraphics[width=0.8\linewidth]{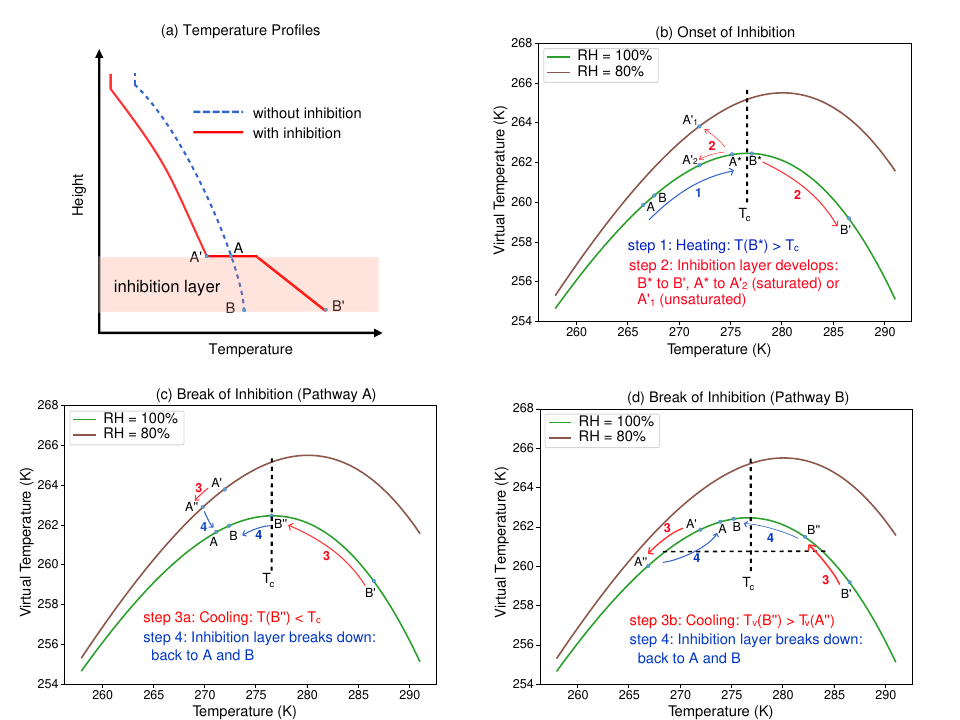}
    \caption{Schematic plots of convective and inhibited states and the transitions between them. \textbf{(a)} Atmospheric temperature profile with or without inhibition. Point A represents the air above the inhibition level; point B represents the inhibition layer. The transition at the inhibition level is approximated as a discontinuity, and the thickness of the inhibition layer is exaggerated. \textbf{(b)-(d)} The evolution of points A and B on the $T$-$T_v$ plot for $\wa$ in a background air with $M_d=2.3\unit{g/mol}$. The virtual temperature $T_v=\frac{TM_d}{\mu}$ ($\mu$ is the average molecular weight that depends on the vapor concentration) is proportional to $1/\rho$ when the pressure is fixed. In these plots, a parcel of air gets hotter going to the right and less dense going up. The two solid lines show $T_v$ under constant relative humidity (RH). The vertical dashed line shows the Guillot threshold temperature $T_c$, which is also the point at which a saturated air parcel reaches its lowest density (peak of the green curve). The horizontal line is a reference for comparing the virtual temperatures of A'' and B''. (b) The onset of inhibition by warming. (c) \textbf{Pathway A:} The inhibition breaks once the near-surface air becomes cooler than the Guillot threshold. This pathway is more likely when the upper atmosphere is dry. (d) \textbf{Pathway B:} The inhibition breaks once near-surface air becomes less dense than the air at the top of inversion layer. This pathway is more likely when the upper atmosphere is close to saturation, and is always preferred over Pathway A when assuming 100\% relative humidity.}
    \label{fig:cbdplots}
\end{figure}

What happens once the near-surface atmosphere becomes inhibited? Qualitatively, the atmosphere's temperature structure must change as a result of changes in energy fluxes. The surface temperature must increase because without moist convection the surface becomes much less efficient at emitting heat. To compensate for the convective heat flux loss, the temperature gradient in the near-surface inhibition layer grows and becomes superadiabatic.
The atmosphere thus develops a layered structure as shown in Figure \ref{fig:cbdplots}(a), which mimics that found on inhibited gas giants \citep{innes_runaway_2023,leconte_condensation-inhibited_2017}. The blue line in Figure \ref{fig:cbdplots}(a) shows the cool and convective state (a troposphere in radiative-convective equilibrium), while the red line shows the warmed and inhibited state (a convective troposphere overlying a near-surface inhibition layer).

A concise way to follow the transition from convective to inhibited is to consider the virtual temperature $T_v$ above the inhibition layer versus in the near-surface layer. We denote these as points A and B, highlighted in Figure \ref{fig:cbdplots}(a). Here, $T_v=\frac{TM_d}{\mu}$ and $\mu$ is the average molecular weight which depends on the vapor concentration. The virtual temperature is a useful measure of atmospheric density and buoyancy, since $T_v \propto 1/\rho$ (keeping the pressure fixed). If A and B are close enough in altitude so that we can ignore their pressure difference, then A has a lower density $\rho$ than B as long as $T_v(A)>T_v(B)$.

As the hycean atmosphere warms up, points A and B evolve as follows:
\begin{itemize}
\item \textbf{Step 1: Heating} (Fig.~\ref{fig:cbdplots}b, $A/B\rightarrow A^*/B^*$). Assuming a saturated atmosphere, A and B's virtual temperatures move along the green curve shown in Figure \ref{fig:cbdplots}(b). The peak of the green curve is equal to the Guillot threshold. Once B crosses this peak, $T(B') > T_c$, the right-hand side of the green curve turns down and the near-surface air at B becomes denser than the air at A, $T_v(A')>T_v(B')$. Moist convection is now inhibited.

\item \textbf{Step 2: Inhibition Layer Grows} (Fig.~\ref{fig:cbdplots}b, $A^*/B^*\rightarrow A'/B'$): The near-surface air (point B) becomes warmer and denser as it moves toward the bottom right of the plot. Meanwhile, to conserve the planet's OLR, the upper atmosphere (point A) has to cool. The precise evolution of point A depends on the relative humidity of the upper atmosphere and results in two possibilities, shown as $A'_1$ and $A'_2$. If the inhibition layer is effective in suppressing the upward flux of water vapor from the surface to the troposphere, the relative humidity in the troposphere above the inhibition layer drops. In this case, $A'_1$ cools along a subsaturated $T_v$ curve; for illustration purposes, Figure \ref{fig:cbdplots}(b) shows $T_v$ in brown for a relative humidity (RH) value of 80\%. However, if the troposphere remains fully saturated because the upward flux of water vapor remains efficient (e.g., via efficient water vapor diffusion through the inhibition layer), then the upper atmosphere cools following the green curve and the point $A^*$ evolves into $A'_2$.
\end{itemize}

What happens once we reverse warming and cool down the inhibited atmosphere? There are two possibilities, depending on whether the upper atmosphere (point A) is subsaturated or saturated; Figure \ref{fig:cbdplots} shows the corresponding two pathways.
\begin{itemize}
\item \textbf{Step 3a: Cooling in Subsaturated Atmosphere} (Fig.~\ref{fig:cbdplots}c, $A'/B'\rightarrow A''/B''$): As the atmosphere cools, $A'$ evolves into $A''$ along the brown curve, while $B'$ evolves into $B''$ along the green curve. As long as the virtual temperature of $A''$ on the brown curve is higher than the virtual temperature of $B''$ on the green curve, the inhibition layer is stable against large-scale perturbations in which one tries to mix air between the top and bottom of the inhibition layer. However, the air at the bottom of the inhibition layer becomes unstable to small-scale perturbations once the surface temperature drops below the Guillot threshold,
\begin{equation}
    T(B'') < T_c.
\end{equation}
We refer to this mechanism of destabilizing the inhibition layer as Pathway A. 

\item \textbf{Step 3b: Cooling in (almost) Saturated Atmosphere} (Fig.~\ref{fig:cbdplots}d, $A'/B'\rightarrow A''/B''$): Both A and B cool along the green curve. The inhibition layer breaks once the virtual temperature of B on the right side of the green curve moves above that of A on the left side, 
\begin{equation}
    T_v(B'') > T_v(A'').
\end{equation}
Notice that air near the bottom of the inhibition layer is still stable against small-scale perturbations because its temperature exceeds the Guillot threshold ($T(B'')>T_c$). However, the inversion layer becomes unstable to large-scale mixing because air from the bottom of the inversion layer is now less dense than air from the top of the inversion layer. As a result, convection destroys the inversion layer from within.
We refer to this mechanism of destabilizing the inhibition layer as Pathway B.
Our distinction between small-scale versus large-scale instability for Pathway A versus Pathway B matches the numerical results of \citet{habib_3d_2024}, who found that convective adjustment depends on the initial size of the perturbation.

\item \textbf{Step 4: Return to Fully-Convective} ($A''/B''\rightarrow A/B$): Both A and B move back to their original positions.
\end{itemize}

\begin{figure}
    \centering
    \includegraphics[width=0.8\linewidth]{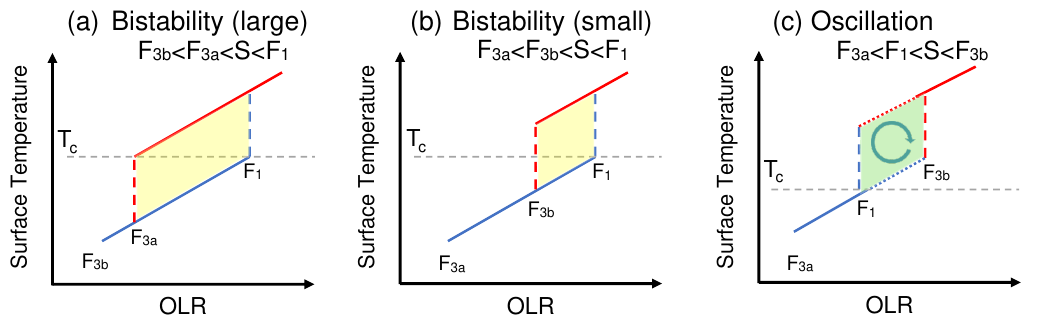}
    \caption{Bistability or oscillations arise because there are three possible relationships between the OLR above which convection breaks down ($F_1$) versus the OLRs below which inhibition breaks down (Pathway A $F_{3a}$, Pathway B $F_{3b}$). If $\max(F_{3a}, F_{3b})<F_1$, both convective and inhibited states are stable, leading to bistability for intermediate values of the instellation $S$, namely $\max(F_{3a}, F_{3b}) < S < F_1$ (yellow shaded region in a,b). \textbf{(a)} When $F_{3b} < F_{3a}  < F_1$, Pathway A is favored. Both edges of the bistable regime satisfy the condition that surface temperature equals the Guillot threshold, $T_s=T_c$, leading to a large bistable regime. \textbf{(b)} When $F_{3a} < F_{3b}  < F_1$, Pathway B is favored. In this case inhibition can break even when the surface temperature is still higher than the Guillot threshold, leading to a smaller bistable regime. This scenario is more likely when the upper atmosphere is close to saturated. \textbf{(c)} When $F_{3a} < F_{1}  < F_{3b}$, neither convective nor inhibited states are stable (represented by dotted lines), so the climate must oscillate between them.}
    \label{fig:Schema_SvT}
\end{figure}

Importantly, the transition from inhibited to convective is not simply the reverse of convective to inhibited because the planet's atmospheric structure is different for each transition, which also affects its OLR. Let's denote the planet's OLR at steps 1, 3a and 3b in Figure~\ref{fig:cbdplots} as $F_1$, $F_{3a}$ and $F_{3b}$. If the planet starts in a convective state, the instellation $S$ has to be higher than $F_1$ to push the planet into an inhibited state. However, if the planet starts in an inhibited state, the instellation has to be lower than $F_3=\max(F_{3a}, F_{3b})$ to drag the planet back into a convective state. 
Therefore, there are four possible equilibrium states. If $S$ is higher than both $F_1$ and $F_3$, the planet can only be in equilibrium in the inhibited regime; if $S$ is lower than both $F_1$ and $F_3$, the planet can only be in equilibrium in the convective regime. At intermediate instellation there are two novel climate regimes, as shown in Figure~\ref{fig:Schema_SvT}:
\begin{itemize}
    \item \textbf{Bistability}: $F_3 < S < F_1$. No matter whether the planet starts in the convective or inhibited state, the initial state remains stable. If the inhibition layer is strong enough to suppress most upward latent heat flux, the upper atmosphere will be depleted in water, favoring Pathway A over Pathway B ($F_{3a}>F_{3b}$). This results in the situation shown in Figure~\ref{fig:Schema_SvT}(a), and the bistable regime is large. If water vapor can penetrate the inhibition layer effectively, the upper atmosphere will be wetter and denser, favoring Pathway B ($F_{3a} < F_{3b}$). This results in a smaller bistable regime (Figure~\ref{fig:Schema_SvT}(b)).
    \item \textbf{Oscillations}: $F_1 < S < F_3$. Since $S$ is both higher than $F_1$ and lower than $F_{3b}$ at the same time, neither state can be in stable TOA energy balance (Figure~\ref{fig:Schema_SvT}(c)). Therefore, the planet must oscillate between convective and inhibited states. Oscillations are more likely to occur when the inhibition is relatively weak and easy to break.
\end{itemize}

\subsection{A Grey Estimate for Bistable versus Oscillating Regimes}\label{sec:regimes}

Our argument above predicts four possible climate regimes. In what part of the parameter space can these regimes be found? To address this question, we develop a semi-analytical model based on grey radiative transfer and use it to estimate the instellation thresholds that correspond to the different climate transitions outlined above (convection breaks, or inhibition breaks via Pathway A and B). Based on these thresholds, we locate the four climate regimes in a 2D parameter space of instellation and surface pressure.

To find the flux threshold at which convection breaks down, we assume an all-troposphere atmosphere that is saturated (RH=100\%). We further assume that the optical thickness of the atmosphere is only a function of pressure, following $\tau\propto p^n$. This ignores the small impact of the purely-radiative stratosphere, and further ignores the variable greenhouse effect of H$_2$O. The temperature profile is a dry or moist adiabat, which we idealize as $T\propto p^\beta$ following \citet{robinson_analytic_2012}, where $\beta$ is constant for simplicity. Top-of-atmosphere energy balance requires that the net incoming stellar flux $S$ equals the outgoing longwave radiation (OLR), so
\begin{equation}\label{eq:enter}
    S=\sigma T_s^4 e^{-D\tau_s} + D\sigma T^4_a\int_0^{\tau_s}\left(\tau/\tau_s\right)^{-4\beta/n}e^{-D\tau}\dd\tau,
\end{equation}
where $\sigma$ is the Stefan-Boltzmann constant, $D$ is the diffusive angular distribution factor in the two-stream radiative transfer equations \citep{raymond_t_pierrehumbert_principles_2010}, and $\tau_s$ is the column-integrated optical thickness. $T_s$ and $T_a$ are the temperature of the surface and near-surface air, respectively, which in the convective state are roughly the same, $T_s\approx T_a$. Note that Equation \ref{eq:enter} can be applied to compute the OLR and instellation of any convective state as a function of $T_s$. To find the instellation at which convection breaks down, we add the constraint that the near-surface and surface temperatures must be equal to the Guillot threshold from Equation \ref{eq:Tc}, $T_s=T_a=T_c(p_s)$. This yields a semi-analytic estimate for the threshold instellation $S=F_1$ at which the convective state becomes inhibited.

To find the flux threshold at which inhibition breaks down, one needs to distinguish between pathways A and B in Figure \ref{fig:cbdplots}. In both cases, we approximate the inhibition layer as a temperature discontinuity right above the surface. This assumption is reasonable since we focus here on scenarios near the onset of convective inhibition, so the inhibition layer is very shallow. We still use Equation \ref{eq:enter} for the top-of-atmosphere energy balance, the only difference is that $T_s$ and $T_a$ are not equal in the inhibited state. To constrain the magnitude of the temperature jump between $T_s$ and $T_a$, we use the surface's energy balance. Since convection is inhibited, fluxes in and out of the surface are purely radiative,
\begin{equation}\label{eq:surf_energy}
    Se^{-k\tau_s} + D\sigma T^4_a\int_0^{\tau_s}\left(\tau/\tau_s\right)^{-4\beta/n}e^{D(\tau-\tau_s)}\dd\tau=\sigma T_s^4,
\end{equation}
where $k$ is an attenuation factor that captures how strongly the atmosphere absorbs stellar radiation and reduces the shortwave flux that reaches the surface \citep{robinson_analytic_2012}. For atmospheres that break down via Pathway A, the corresponding flux threshold $S=F_{3a}$ can be obtained by simultaneously solving Equations \ref{eq:enter} and \ref{eq:surf_energy} together with the Guillot threshold $T_s=T_c(p_s)$. For atmospheres that break down via Pathway B, the flux threshold is $S=F_{3b}$ and one has to replace the condition $T_s=T_c(p_s)$ with $T_{v}(T_a)=T_{v}(T_s)$.

Which of the two pathways is actually realized depends on the relative humidity of the upper atmosphere in the inhibited state, so we distinguish between saturated and subsaturated atmospheres. For a saturated atmosphere, our previous reasoning showed Pathway B is more likely (see Figure \ref{fig:Schema_SvT}(d)). In this case we therefore assume breakdown always follows Pathway B, and the upper atmosphere follows a moist adiabat ($\beta=\beta_{moist}$).
For a subsaturated atmosphere, we do not know in advance which pathway causes the inhibition layer to break down. We therefore calculate both thresholds assuming RH=80\% and choose the larger one, $S=\max(F_{3a},F_{3b})$, plus assume that the atmosphere follows a dry adiabat ($\beta=\beta_{dry}>\beta_{moist}$). It turns out that the assumed relative humidity value only slightly changes the threshold for a subsaturated atmosphere at low surface pressure ($p_s<1\unit{bar}$).
We solve the above equations for surface pressures ranging from $0.1$ to $3\unit{bar}$, using the parameters listed in Table \ref{tab:Ana}.

\begin{figure}[t]
    \centering
    \includegraphics[width=0.8\linewidth]{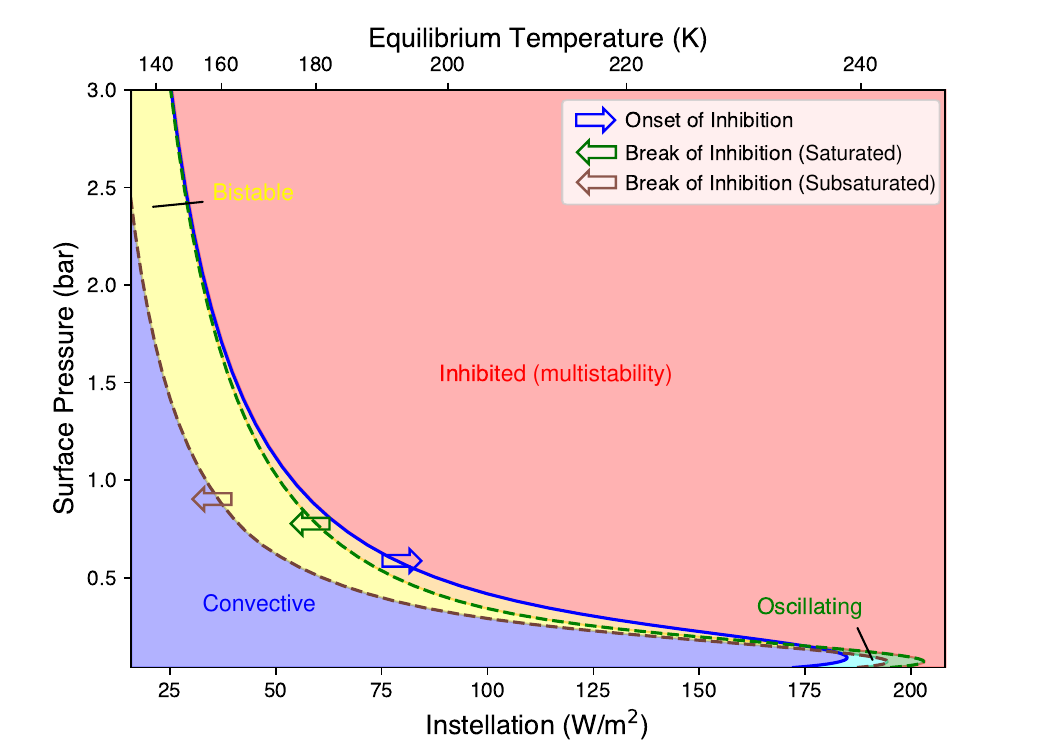}
    \caption{Regime diagram showing four distinct climate regimes, according to our semi-analytical model. The solid blue line represents the instellation above which an atmosphere enters an inhibited state, dashed lines represent the instellation below which an atmosphere becomes fully convective (green versus brown assumes the atmosphere above the inhibition layer follows a moist versus dry adiabat). Arrows indicate the direction of transition.  
    In the lower left of parameter space, atmospheres are convective (blue), in the upper right they are inhibited (red); note, atmospheres in the inhibited regime additionally show multistability (see Section \ref{sec:sim_rslt}). Between the convective and the inhibited regimes, thicker atmospheres are bistable (light and dark yellow) and thinner atmospheres oscillate (light and dark green). All parameters used in the semi-analytical model are listed in Table \ref{tab:Ana}.}
    \label{fig:AnaRegimes}
\end{figure}

Figure \ref{fig:AnaRegimes} shows the results of our semi-analytical model. The solid blue line shows the instellation at which convection becomes inhibited under warming (going from left to right), and the dashed lines (green for the saturated case and brown for the subsaturated case) show the instellation at which inhibition breaks down under cooling (going from right to left).
These lines divide the parameter space into four possible regimes: in the red region to the right of both solid and dashed lines, an atmosphere is always inhibited; in the blue region to the left of all lines, it is always convective. The yellow region enclosed by the dashed line on the left and the solid line on the right represents the bistable regime, whereas the green region enclosed by the solid line on the left and the dashed line on the right represents the oscillating regime.

\begin{table}[]
    \centering
    \begin{tabular}{l c c}
    \toprule
    & Parameter & Value \\
    \midrule
 Reference optical thickness &$\tau_0(1\unit{bar})$&$20$\\
         Optical thickness gradient &$n=\ddx{\ln \tau}{\ln p}$&  $2$\\
         Shortwave attenuation factor &$k=\frac{\tau_{SW}}{\tau_{LW}}$& $0.01$\\
 Diffusive factor in longwave radiation &$D$&$2$\\
 Adiabatic gradient &$\beta_{moist}\ (\beta_{dry})$&$0.23\ (2/7)$\\
 Molecular weight &$M_d\ (M_v)$&$2.3\ (18)\unit{g/mol}$\\
 Gravity &$g$&$10\unit{m/s^2}$\\
 \bottomrule
 \end{tabular}
    \caption{Parameters used in our semi-analytical model.}
    \label{tab:Ana}
\end{table}

Based on Figure \ref{fig:AnaRegimes}, we expect that hycean planets exhibit bistability over a considerable part of parameter space, especially if the upper atmosphere in the inhibited regime becomes subsaturated (i.e., if the inhibition layer is effective in blocking upward water vapor transport, thus depleting the upper atmosphere of water).
For thin atmospheres with surface pressures $\mathcal{O}(0.1)\unit{bar}$, we expect oscillations. For atmospheres thicker than $\mathcal{O}(0.1)\unit{bar}$, we expect bistability.
Note that once an inhibited atmosphere becomes much warmer than the Guillot threshold, the approximation that the inhibition layer is shallow starts to break. As we discuss in Section \ref{sec:sim_rslt}, this can lead to multistability in which the thickness of the inhibition layer as well as the planet's surface temperature decouple from the top-of-atmosphere energy budget and become poorly constrained. To test these theoretical predictions, we use a 1D numerical model, which is described in the next section.

\section{Numerical 1D Model}\label{1dmodel}

We use a 1D radiative-convective model to test and verify our theoretical predictions. The model uses time-stepping to simulate the evolution of a 1D atmospheric column coupled to a slab ocean surface. Important model components are its non-grey radiative transfer scheme, for which we use \texttt{exo-k} \citep{leconte_spectral_2021}, and a generalized moist convection scheme valid for arbitrary atmospheric compositions \citep{ding_convection_2016}. Precipitation is instantly removed (see below), so the atmosphere is always cloud-free.

\subsection{Thermodynamics and Other Settings}
We ignore the critical behavior of water vapor because the temperature range considered is much lower than the critical temperature of water ($647\unit{K}$). The heat capacity of air is assumed to be constant, so the latent heat of vaporization changes linearly with temperature. Once a layer reaches supersaturation, we assume instantaneous condensation and precipitation. This is a good approximation for the time scales ($\sim1$ days or longer) that we are interested in. Reevaporation of precipitation is ignored, and the energy and mass associated with precipitation are deposited in the surface. The potential influence of reevaporation is discussed in Section \ref{sec:shortcoming}.

The model is unevenly divided into 80 layers using a sigma coordinate with higher resolution near the surface and a model top at $\sim 1\unit{Pascal}$. For simulations with an inhibition layer, changes in model resolution can affect temperature of the lower atmosphere and surface, since temperature gradients are very large in this region. The acceleration of gravity is set to $9.8\unit{m/s^2}$, and the consequences of changing this value are discussed in Section \ref{sec:dis}.

\subsection{Radiative Transfer}
The non-grey radiative transfer is performed using \texttt{exo-k} version 1.2.2 \citep{leconte_spectral_2021}. 
We started this work before \texttt{exo-k} was fully capable of dealing with convective inhibition, so we only use \texttt{exo-k} to calculate radiative fluxes at each timestep and ignore its more recently added evolution utilities \citep{leconte_3d_2024}.

The atmosphere consists of $\hh$, $\mathrm{He}$, and $\wa$. The mass fraction of $\mathrm{He}$ in dry air is 0.25188, which is approximately the solar value \citep{asplund_chemical_2009} and is the same as that used by \cite{innes_runaway_2023}. We use $\wa$ cross section data from \texttt{TauREx} between 0.3 to 50 $\mathrm{\mu m}$, which are publicly available via the ExoMOL website \citep{polyansky_exomol_2018, chubb_exomolop_2021}, and $\hh$ and $\mathrm{He}$ collision-induced absorption data from \texttt{HITRAN-2011} \citep{karman_update_2019}. The stellar spectrum is a $5570~\mathrm{K}$ blackbody truncated to the same wavelength range. The surface albedo is taken to be $0.12$.

\subsection{Convection and Surface Physics}

Convection is simulated using the instantaneous moist convection adjustment, or Manabe-type, scheme from \cite{ding_convection_2016}. To make the scheme compatible with convective inhibition, we introduce the virtual effect into the criteria for convective stability. 
At each timestep, we compare the potential temperature (for unsaturated layers) or virtual potential temperature (for saturated layers) of adjacent layers and perform convective adjustment on every unstable pair. If both layers are initially unsaturated, they will be adjusted to a dry adiabat with a uniform $\wa$ mixing ratio. If any of the layers are saturated, they will be adjusted to a moist adiabat, and the final mixing ratios are determined following the method suggested by \cite{ding_convection_2016}.

Surface heat fluxes are parameterized using standard bulk formulae \citep{raymond_t_pierrehumbert_principles_2010}.
The energy equation at the surface is:
\begin{equation}\label{Eq:surf}
\rho_w h c_{pl} \frac{dT_s}{dt} = F_{net} + P(T_{rain}-T_s)c_{pl} - E\left[ L(T_s) + (T_a-T_s)c_{pv}\right] - c_1 w \rho_a c_{pa}(T_s-T_a)+F_{other},
\end{equation}
where $\rho_w$ and $\rho_a$ are the density of liquid water and air, $h$ is the depth of the ocean, $c_{pl}$, $c_{pv}$, and $c_{pa}$ are the specific heat capacities of liquid water, water vapor, and air, $F_{net}$ is the net downward radiative flux, $P$ is the precipitation rate, $E = wc_2(q_{sat}(T_s) - q_a)$ is the evaporation rate, $L$ is the specific latent heat of water, $w$ is the surface wind speed, $q_{sat}$ is the saturation water mixing ratio, and $q_a$ is the mixing ratio of water vapor in the near-surface air, $c_1$ and $c_2$ are the sensible heat flux coefficient and the evaporation rate coefficient, $T_{rain}$, $T_{s}$, and $T_{a}$ are the temperatures of the rain, the surface, and the near-surface air, and $F_{other}$ contains the potential energy of the precipitation. The first four terms on the right-hand side of Equation \ref{Eq:surf}  represent the heat fluxes carried by radiation, precipitation, evaporation, and sensible heat, respectively. Since we do not care about changes in ocean level, $h$ is set to a constant value.

We set the parameters $c_1$, $c_2$, $w$, and $h$ equal to $10^{-3}$, $10^{-6}$, $10\unit{m/s}$, and $10\unit{m}$, respectively. In most cases, changing these parameters only has a slight effect on the equilibrium state. However, for cases in the oscillating regime, they affect the oscillation period (discussed in Section \ref{sec:con}). In these cases, we change $w$ to $1\unit{m/s}$, which ensures that each phase is longer than the typical interval over which model outputs are saved.

\subsection{Moisture Diffusion}
\label{Sec:diffS}

Sub-scale mixing of moisture is parameterized as vertical diffusion of water vapor,
\begin{equation} \label{eq:diff}
    \frac{\partial x}{\partial t} = K_{zz}\frac{\partial^2 x}{\partial z^2},
\end{equation}
where $x$ is the mole mixing ratio of water vapor and $K_{zz}$ is the eddy diffusivity, which we assume is vertically uniform.
The resulting moisture flux matters little in convective parts of the atmosphere, because for the $K_{zz}$ values we consider moist convection always transports moisture much more efficiently than diffusion. However, it turns out to be very important in the inhibition layer, because in this case $K_{zz}$ determines the upward flux of water and latent heat through those parts of the atmosphere in which convection is suppressed.

To constrain the vertical diffusivity of water vapor, we consider the following limits. In the limit of a stagnant inhibition layer, in which turbulence is completely suppressed, diffusion occurs only via molecular diffusion. The diffusion coefficient of $\wa$ in $\hh$ is on the order of $10^{-4} \unit{m^2/s}$ at $p\sim1\unit{bar}$ \citep{schwertz_diffusivity_1951}.

In the opposite extreme, the inhibition layer might be efficiently stirred by turbulence. For example, even if the inhibition layer itself is stagnant, reevaporation of precipitation in the overlying troposphere might lead to convective downdrafts that penetrate into the near-surface inhibition layer and generate turbulent mixing. Similarly, horizontal temperature gradients and a large-scale atmospheric circulation could result in shear, which can generate mechanical turbulence. In either case, the mixing of water vapor through the inhibition layer would be significantly increased compared to the molecular diffusion limit. 3D simulations of hycean atmospheres by \cite{leconte_3d_2024} suggest that the eddy diffusivity in the inhibition layer is on the order of $0.1$ to $1\unit{m^2/s}$. Because the simulations in \cite{leconte_3d_2024} did not consider the potential additional mixing due to a large-scale atmospheric circulation, we set our upper limit for $K_{zz}$ one order of magnitude higher, to $20\unit{m^2/s}$.

\subsection{Simulation Procedure} \label{simu}

To verify our theoretical results, we perform two series of 1D simulations. We consider cool-start and hot-start simulations, for dry air masses ranging from $1\times10^3\unit{kg/m^2}$ to $3\times10^4\unit{kg/m^2}$.

Cool-start and hot-start simulations differ as follows. In the cool-start series we start from a low stellar flux, run the model to equilibrium, then increase stellar flux by a small amount and find a new equilibrium. Here ``low" means that we never find an inhibition layer in any simulations run with the initial stellar flux. After a simulation reaches equilibrium, by which we mean less than $0.02\unit{W/m^2}$ TOA net flux imbalance, the instellation is increased by $2\unit{W/m^2}$ for the next simulation. The next simulation uses the final state of the previous simulation as its initial condition. This procedure is repeated until an inhibition layer appears and, if possible, occupies more than one vertical layer.
In the hot-start series we reverse the above procedure and end simulations once the inhibition layer disappears. In both series, for the initial condition of the first simulation we adopt a moist adiabatic profile (even though a moist adiabat is not close to equilibrium in the hot-start cases).

Because cool-start and hot-start simulations have different initial surface temperatures, they should also have different initial surface pressures. We adjust the initial surface pressure so that, for two series with otherwise-identical parameters, dry air mass is identical.

\begin{figure}
    \centering
    \includegraphics[width=0.8\linewidth]{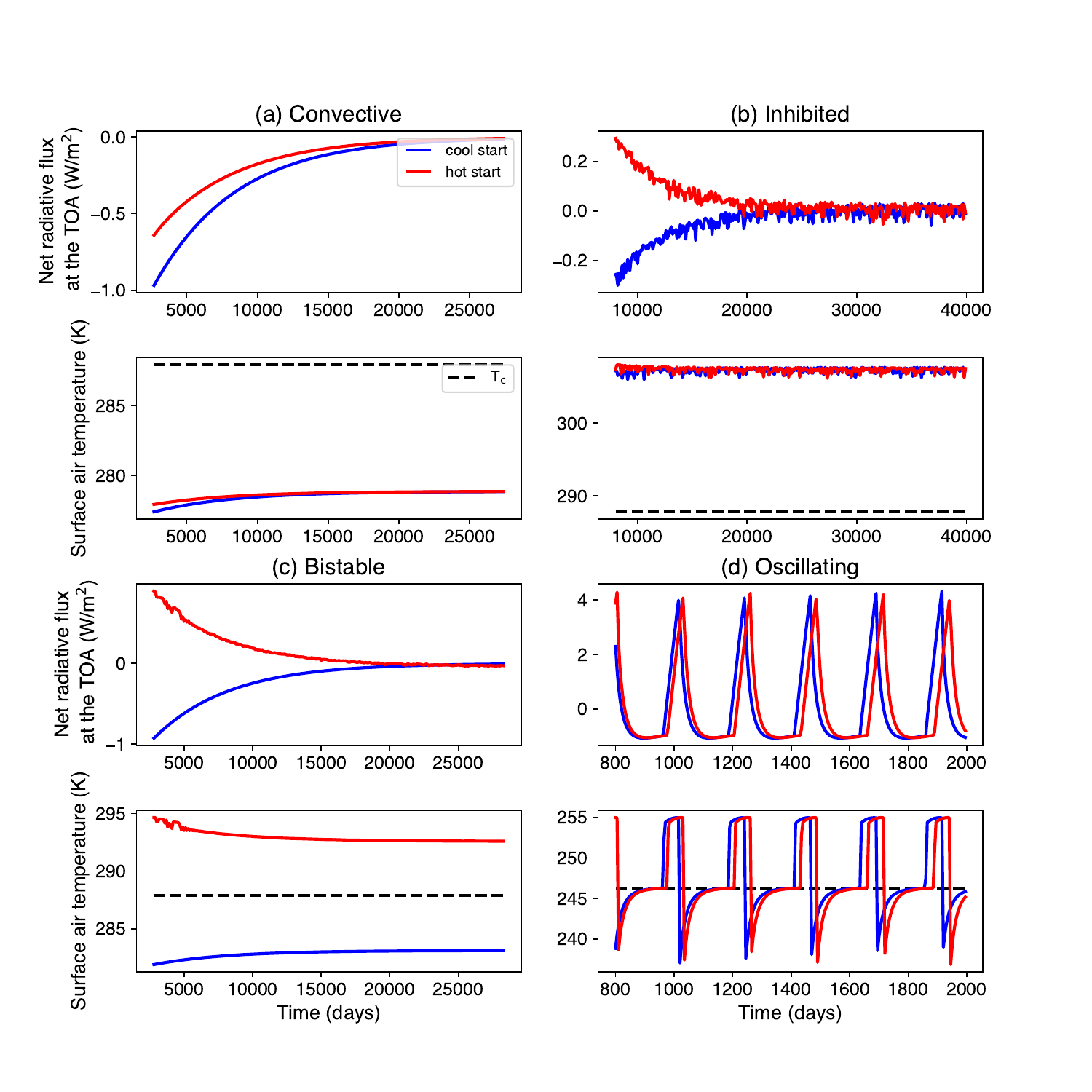}
    \caption{Numerical simulations confirm hycean planets can occupy four different climate regimes. Here we show time series from typical cases in each regime. Dashed black lines show the Guillot threshold temperature $T_c$. Simulation parameters are: $K_{zz}=10^{-4}\unit{m^2/s}$; $p_s=2\unit{bar}$ (a-c) or $p_s=0.1\unit{bar}$ (d); $S=44\unit{W/m^2}$ (a), $S=58\unit{W/m^2}$ (b), $S=48\unit{W/m^2}$ (c), $S=189\unit{W/m^2}$ (d). Because the surface layer is always saturated in these cases, whether the atmosphere convective or inhibited is determined by whether the surface air temperature exceeds $T_c(p_s)$. (a): Both hot-start and cool-start initial conditions lead to a convective state. (b): Both initial conditions lead to an inhibited state. (c): Hot-start leads to an inhibited state, while cool-start leads to a convective state. (d): Both initial conditions lead to an oscillating state.}
    \label{fig:TS-cbd}
\end{figure}

\section{Numerical Results}\label{sec:sim_rslt}

\subsection{Confirmation of Bistability and Oscillations}\label{sec:con}

First, we confirm that the numerical model exhibits four distinct regimes. Figure \ref{fig:TS-cbd} shows timeseries of TOA radiative balance and surface temperature of four representative cases. Each case shows both cool-start and hot-start initial conditions (blue and red lines). To highlight the behavior of each case with regards to the Guillot threshold, the surface temperature plots additionally show $T_c(p_s)$ using horizontal black lines. Detailed simulation parameters are listed in the figure caption.

Figures \ref{fig:TS-cbd} (a) and (b) illustrate the convective and inhibited regimes. Starting from two different initial conditions, both regimes converge to a single equilibrium. In the convective state the equilibrium is a relatively cool surface with $T_s<T_c(p_s)$, in the inhibited state it is a relatively warm surface with $T_s>T_c(p_s)$.
Note that in the convective regime the timeseries are smooth, whereas in the inhibited regime they show small-scale noise. We believe this is due to the model's finite vertical resolution; once the near-surface inhibition layer appears, it is typically resolved by only one to a few vertical grid points near the model's surface. Heating or cooling then causes relatively abrupt jumps in inhibition layer thickness, which is further amplified by the response and feedback of the overlying convective atmosphere.
Both regimes in Figure \ref{fig:TS-cbd} (a) and (b) are compatible with previous theoretical work on convective inhibition \citep{guillot_condensation_1995,leconte_condensation-inhibited_2017}.

As the first novel dynamical regime, Figure \ref{fig:TS-cbd} (c) illustrates a bistable case. Starting from two different initial conditions the simulations converge in two distinct equilibrium states, with a final $T_s$ value above versus below $T_c$. Although both simulations are equilibrated based on their TOA energy budget, their surface temperatures differ by more than $10\unit{K}$.

As the second novel dynamical regime, Figure \ref{fig:TS-cbd} (d) illustrates an oscillating case. Both initial conditions end up in essentially the same cycle. The atmosphere spends a relatively long time in the convective state, warming up while $T_s<T_c(p_s)$. Once the surface air temperature reaches the Guillot threshold, an inhibition layer develops. This suppresses latent heat transport and the surface temperature rapidly jumps up to $T_s>T_c(p_s)$. In the inhibited phase of the cycle, the TOA energy imbalance then turns positive, and the climate starts to cool. At some point, the inhibition becomes unstable and disappears. The surface temperature falls again below the Guillot threshold, $T_s<T_c(p_s)$, and the cycle repeats. 

In the oscillating regime, the property of the surface can affect length of the cycle. In the limit of high surface heat capacity (e.g., if the mixed layer is thick), the surface can be treated as an isothermal heat reservoir whose temperature exceeds the Guillot threshold. This can lead to a transient convective phase if the surface-air heat exchange is efficient, so we deliberately reduce the sensible heat flux coefficient in oscillating cases for better illustration. In the limit of low surface heat capacity (e.g., if the mixed layer is shallow), the cycle length is solely determined by the atmospheric structure in the inhibited state. A thick and hot inhibition layer would lead to a convective phase shorter than the inhibited phase, while a shallow, or even marginal, inhibition layer would lead to a convective phase longer than the inhibited phase.

\subsection{Confirmation of the Predicted Boundary between Convective and Inhibited Regimes}

\begin{figure}
    \centering
    \includegraphics[width=0.8\linewidth]{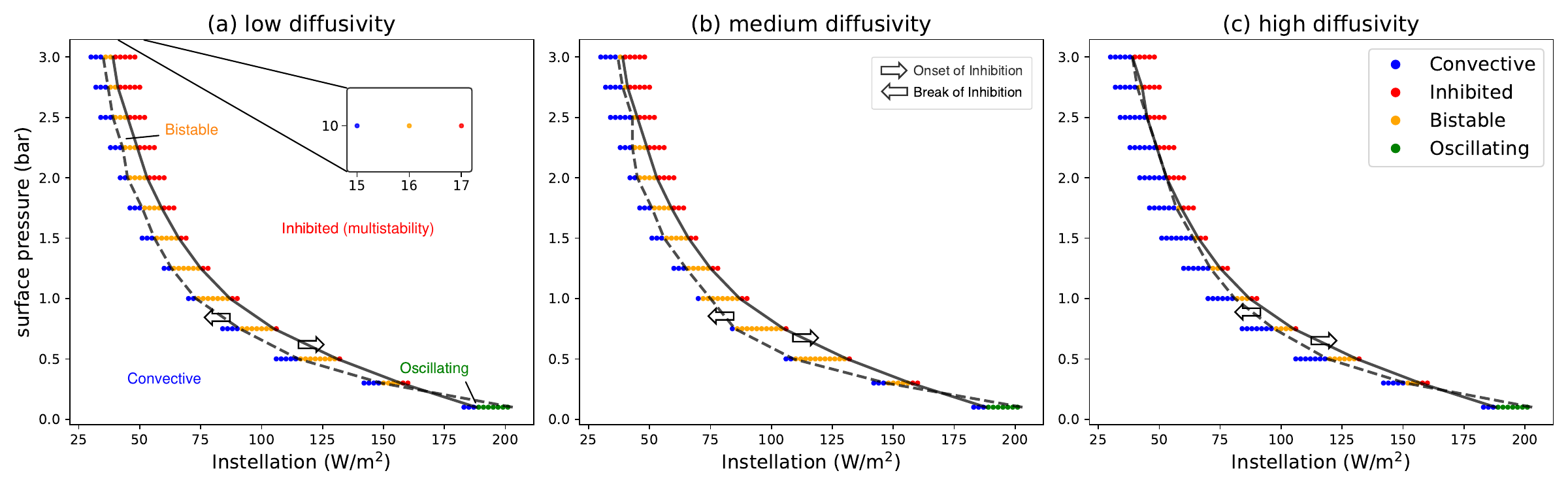}
    \caption{Regime diagram for the four different dynamical regimes, based on numerical 1D simulations. Compare this figure against the prediction from our semi-analytical model (Figure \ref{fig:AnaRegimes}). Left to right panels show results for different diffusivities ($K_{zz}=10^{-4}$, $0.2$, and $2\unit{m^2/s}$, respectively). Blue, yellow, red, and green dots represent convective, inhibited, bistable, and oscillating cases, respectively. The solid line represents the threshold for the onset of inhibition, and the dashed line represents the threshold for the break of inhibition. In agreement with our semi-analytical prediction, there is an oscillating regime for thin atmospheres around $0.1\unit{bar}$ and a considerable bistable regime for atmospheres above $0.3\unit{bar}$.}
    \label{fig:sim_diagram}
\end{figure}

Up to now, we have confirmed that the numerical model exhibits bistable and oscillating states, in qualitative agreement with our theory. In which part of parameter space do these states occur, and does their occurrence match our semi-analytical prediction in Figure \ref{fig:AnaRegimes}?

Figure \ref{fig:sim_diagram} shows the parameter space which we explore numerically. Each combination of instellation and surface pressure, shown in colored points, is explored using a cool-start and a hot-start simulation. The final states of these simulations are categorized into one of four possible states, according to the standard used in Figure \ref{fig:TS-cbd}. Different panels in Figure \ref{fig:sim_diagram} explore the impact of water vapor diffusivity, ranging from low to high $K_{zz}$ values. Note that we do use a non-zero surface albedo (0.12), so the absorbed stellar flux is lower than the value indicated in the plot and the diagram would approximately be shifted horizontally if another surface albedo value is chosen.

Overall, Figure \ref{fig:sim_diagram} agrees well with our analytical prediction in Figure \ref{fig:AnaRegimes}. Moist convection is inhibited at high stellar flux (red points), and the atmosphere is fully convective at low stellar flux (blue points). The transition between these two regimes occupies a diagonal region, ranging from top left to bottom right. Inside the transition region, most cases are bistable (yellow points). Oscillations occur only for cases with thin atmospheres, with surface pressures around $0.1$ bar (green points). As the atmosphere becomes thicker, the width of the bistable regime decreases. However, even at $10$ $\mathrm{bar}$ it is still not negligible and takes up about $2\unit{W/m^2}$ (see Figure \ref{fig:sim_diagram} a, inset). For comparison, because the hycean habitable zone moves away from the star at higher surface pressures \citep{innes_runaway_2023}, having bistability over a region $2\unit{W/m^2}$ wide amounts to a significant fraction of the habitable zone.

Different panels in Figure \ref{fig:sim_diagram} show how our results depend on the water vapor diffusivity $K_{zz}$. In agreement with our theoretical predictions, we find that the width of the bistable region is sensitive to $K_{zz}$. For the low $K_{zz}$ in Figure \ref{fig:sim_diagram}(a), we find that the upper atmospheres of inhibited states are almost completely dry. In contrast, for the high $K_{zz}$ in Figure \ref{fig:sim_diagram}(c), we find that the upper atmospheres of inhibited states are almost completely saturated. As a result, the bistable regime is smaller at higher diffusivity. This matches our theoretical prediction that a drier atmosphere favors Pathway A over Pathway B, which results in a narrower bistable regime (see dashed lines for subsaturated versus saturated in Figure \ref{fig:AnaRegimes}).

Extrapolating this reasoning, one might expect that for medium $K_{zz}$ the width of the bistable regime should also be intermediate. However, Figure \ref{fig:sim_diagram}(b) shows that this is only true for atmospheres thicker than $1.5\unit{bar}$, whereas for thinner atmospheres, the bistable regime is the widest for medium $K_{zz}$. This is because adding water vapor to the atmosphere has two effects. The first is to increase the saturated region above the inhibition layer that follows a moist adiabat instead of a dry one, thus reducing the overall lapse rate, leading to a smaller bistable regime. The other is to act as a greenhouse gas that stabilizes the inhibition layer, leading to a wider bistable regime.
At high surface pressures, the atmosphere's greenhouse effect is dominated by the collision-induced absorption of $\hh$ and $\mathrm{He}$, so the greenhouse effect of $\wa$ is less important and the bistable regime is narrower. At low surface pressures, $\hh$ is no longer an effective greenhouse gas, so the greenhouse effect of $\wa$ is more significant and the bistable regime is wider than expected.

\begin{figure}[h]
    \centering
    \includegraphics[width=0.8\linewidth]{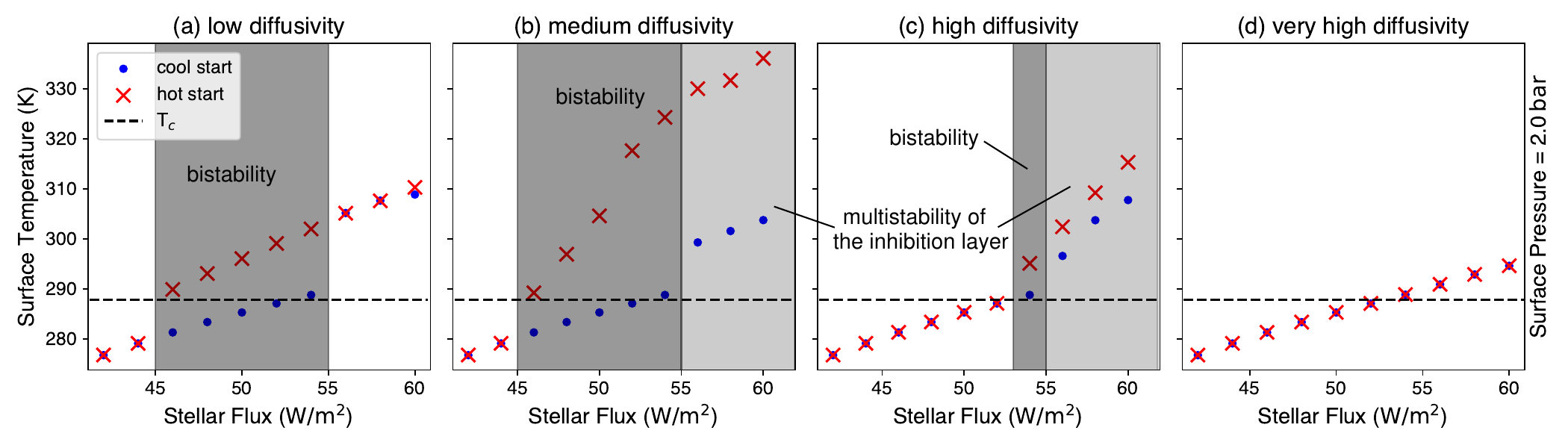}
    \caption{Hysteresis diagram for surface temperature as a function of stellar flux and vertical diffusivity. Here $p_{dry}=2.0\unit{bar}$ and diffusivity ranges from low to very high from left to right ($K_{zz}=10^{-4}$, $0.2$, $2$, and $20\unit{m^2/s}$). Each dot represents a cool-start or a hot-start simulation. The dark-shaded regions represent the bistable regime, with either a thin inhibition layer or no inhibition layer. The light-shaded regions highlight simulations outside the bistable region, but for which cool-start and hot-start simulations still do not converge. This is because, for simulations with thick inhibition layers at high stellar flux, the inhibition layer's thickness becomes poorly constrained and leads to multistability (see Section \ref{sec:multi}).}
    \label{fig:hys}
\end{figure}

What does bistability imply for the climates of hycean planets? One immediate consequence is hysteresis. Figure \ref{fig:hys} shows the equilibrated surface temperatures as a function of stellar flux in cool-start versus hot-start simulations. We plot simulations at moderately high surface pressure ($p_{dry}=2.0\unit{bar}$), to avoid any oscillating cases. 
The hysteresis of surface temperature is clearest in Figure \ref{fig:hys} (a). At low and high stellar fluxes, cool-start and hot-start simulations converge to essentially the same state. However, inside the bistable region highlighted in dark grey, surface temperature depends on the simulation's initial condition. Cool-start cases that started convective and thus have cool surfaces stay cool; hot-start cases that started inhibited and thus have hot surfaces stay hot.
This behavior is robust to large changes in water vapor diffusivity: inside the bistable region, surface temperature always shows strong hysteresis. The magnitude of hysteresis depends on the assumed diffusivity, but in Figure \ref{fig:hys} (b) it reaches up to $40\unit{K}$.
The only exception are simulations with very high diffusivity (see Figure \ref{fig:hys}d), which do not exhibit bistability and therefore show no apparent hysteresis. In this limit, the large diffusivity is always effective in transporting latent heat away from the surface. Even though an inhibition layer still develops, it has only a negligible effect on the atmosphere's temperature-pressure profile, so the bistability associated with the inhibition is also negligible.

Surprisingly, Figure \ref{fig:hys} also shows that some cool-start and hot-start simulations do not necessarily converge at high stellar flux. For example, in Figure \ref{fig:hys} (b), above $S=55\unit{W/m^2}$ the equilibrated surface temperatures of cool-start and hot-start simulations still differ from each by about $30\unit{K}$. Both initial conditions lead to near-surface inhibition layers, yet the inhibition layers are not identical. This stands in contrast to our simple theoretical expectation (Section \ref{sec:theory}) that, at high enough stellar flux, bistability should disappear and all simulations should merge. To explain why, we consider multistability next.

\subsection{Multistability In Hot Atmospheres}\label{sec:multi}

\begin{figure}[h!]
    \centering
    \includegraphics[width=0.8\linewidth]{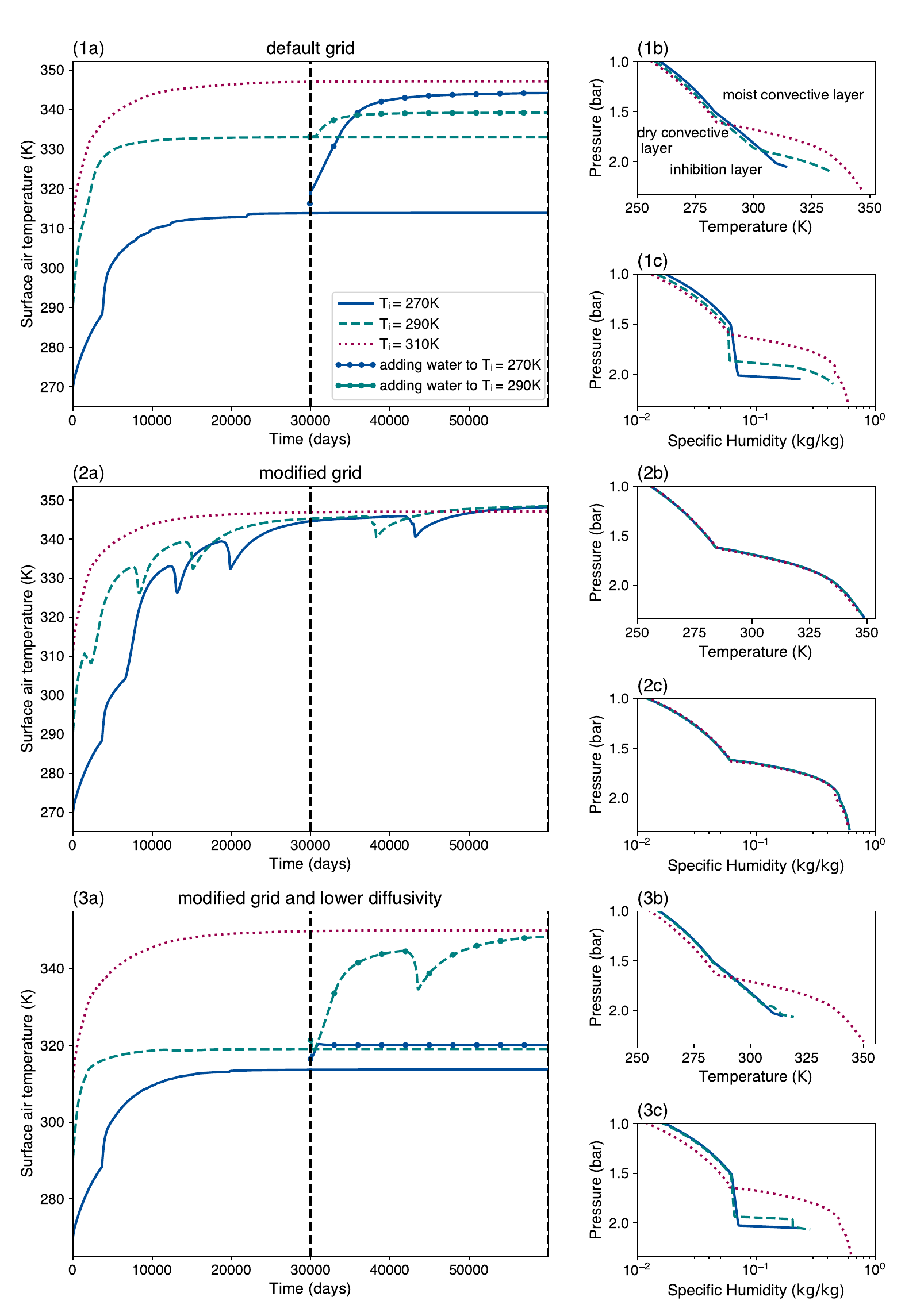}\\
    \caption{Warm atmospheres with thick inhibition layers show multiple stable equilibria, or multistability. For each series of simulations, left panels show time series of surface temperature for three different initial states (moist adiabats with surface temperature $T_i = 270,\ 290,\ 310\unit{K}$), right panels show $T$-$p$ and $q$-$p$ profiles at the end of simulation. Top panels show simulations with default vertical grid. Middle panels show simulations with a modified grid which effectively doubles vertical resolution in the lower atmosphere; the timestep is also halved. Bottom panels show simulations with  modified grid and timestep, plus a lower diffusivity. Other parameters: $S=70\unit{W/m^2}$, $p_{dry}=2.0\unit{bar}$, and $K_{zz}=0.2\unit{m^2/s}$ for top and middle panels $0.15\unit{m^2/s}$ for bottom panels. At $t=30000\unit{d}$, a subset of simulations is perturbed by adding moisture to the dry convective layer above the inhibition layer, which leads to more stable branches.}
    \label{fig:branches}
\end{figure}

Why do cool-start and hot-start simulations sometimes fail to converge at high stellar flux? To explore this behavior, we perform a second set of simulations. We choose a high stellar flux and a moderately thick background atmosphere ($S=70\unit{W/m^2}$, $p_{dry}=2.0\unit{bar}$, and $K_{zz}=0.2\unit{m^2/s}$). This puts the expected surface temperatures well above the Guillot threshold, and ensures that the inhibition layer is thick. We pick three different initial conditions with surface temperatures between $270\unit{K}$ and $310\unit{K}$, and run the model to equilibrium. To explore the effect of perturbations to the model's inhibition layer, at $t=30000\unit{d}$ we add a moisture perturbation to the layer of dry convection above the inhibition layer. In addition, to explore the importance of numerical effects, we repeat the same experiment with two model modifications. In one, we modify the vertical grid to an equal pressure coordinate, which effectively doubles the model's vertical resolution in the lower atmosphere. In the other, we modify the vertical grid in the same way, but additionally reduce the diffusivity to $K_{zz}=0.15\unit{m^2/s}$.

We find that all our default simulations show clear multistability. Figure \ref{fig:branches}(1a) shows the time evolution of the surface air temperature, while Figure \ref{fig:branches}(1b, 1c) show the equilibrated vertical temperature-pressure and specific humidity profiles. Two features stand out. First, hotter initial states develop thicker inhibition layers and equilibrate at higher surface temperatures. Even though all simulations receive the same incoming stellar flux, the one with the coldest initial condition (blue line) equilibrates with a relatively cool surface ($310~\unit{K}$) and a shallow inhibition layer, while the one with the hottest initial condition (red line) equilibrates with a much hotter surface (about $345~\unit{K}$) and an inhibition layer that extends from 2.5 up to 2 bar.
Second, by perturbing the atmosphere's humidity, one can make the atmosphere shift to another equilibrium. In some cases, the perturbation can cause two different states to merge into the same equilibrium; for example, some of the perturbations to light green and blue lines end up in a new state intermediate between green and red. This further suggests some of the multiple equilibria are more stable than others.

How much is multistability affected by numerical details? To explore the question, we redistribute the 1D model's 80 vertical grid points to an equal delta-p grid. Doing so effectively doubles the model's vertical resolution in the lower atmosphere. The timestep is shortened accordingly by half.
Figure \ref{fig:branches}(2a, 2b) show that, in this case, multistability largely disappears and the equilibrated surface states are much more similar than before. Note that even in this case the equilibrated states are not identical. Moreover, the two colder initial conditions (blue and green lines) now display abrupt jumps during the model's spin-up, suggesting that some simulations are briefly attracted by neighboring states before all eventually settle into a similar equilibrium.

Next, we use the modified vertical grid, but also reduce the model's diffusivity. In this case, the model shows multistability again, although some states are more likely to merge than with the default grid (see Figure \ref{fig:branches}(3a)). Using the default grid our 3 initial conditions and 4 perturbations identify about 5 distinct equilibria, whereas with the modified grid and lowered diffusivity these simulations converge to about 4 distinct equilibria. The model's dependency on vertical resolution is likely due to the atmosphere's humidity discontinuity right above the inhibition layer (see Figure \ref{fig:branches}(1c, 3c)). This strong moisture gradient makes the results sensitive to grid thickness and diffusion efficiency, such that higher vertical resolution can be compensated via lower diffusivity.

Multistability thus appears to be a robust feature of hycean atmospheres with thick inhibition layers. Theoretically, one might expect such behavior. In radiative-convective equilibrium the atmosphere's water cycle is closely tied to moist convection. In contrast, on hycean planets radiative-convective equilibrium breaks down near the surface and the water cycle provides an additional degree of freedom.
To uniquely constrain this degree of freedom, one needs additional constraints. For example, \citet{innes_runaway_2023} assumed that the atmosphere is always saturated. Their results thus likely resemble our simulations in the limit in which the inhibition layer is thick and the lower atmosphere very humid. However, we see no reason to prefer this limit over the possible alternatives. As our simulations show, for the same instellation, the atmosphere could also support multiple alternative states with drier lower atmospheres, thinner inhibition layers, and colder surfaces.

\subsection{How does Diffusivity Affect Hot Atmospheres}\label{sec:diff}

\begin{figure}
    \centering
    \includegraphics[width=0.8\linewidth]{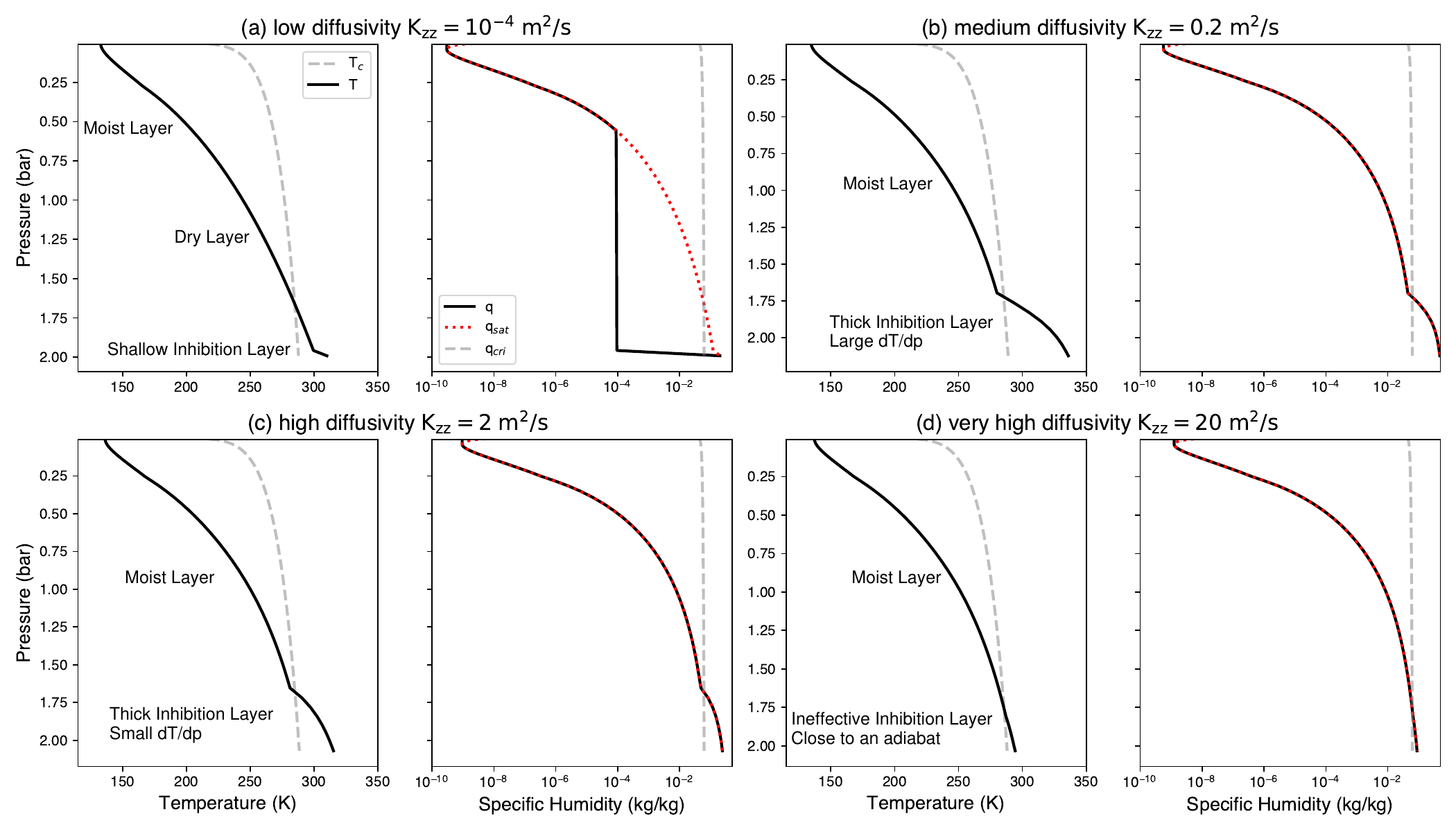}
    \caption{For hot atmospheres in the inhibited regime, the near-surface climate is highly sensitive to the assumed water vapor diffusivity through the inhibition layer. Left sub-panels show temperature-pressure, right sub-panels specific humidity-pressure profiles. Simulations shown here use $p_{dry}=2\unit{bar}$, $S=60\unit{W/m^2}$ and start with a hot initial condition, but differ in terms of $K_{zz}$. In all cases, the surface is hot enough to lie above the Guillot threshold ($T_s>T_c$). The intersections of $p\text{-}q$ and $p\text{-}q_{cri}$ (or equivalently, the intersections of $p\text{-}T$ and $p\text{-}T_c$ in saturated cases) mark the top of the inhibition layer.}
    \label{fig:diff_comp}
\end{figure}

So far we have confirmed our theoretical prediction that hycean atmospheres close to the Guillot threshold exhibit bistability and oscillations (Figure~\ref{fig:TS-cbd}). In addition, we found that hycean atmospheres far above the Guillot threshold exhibit multiple stable equilibria (Figures~\ref{fig:hys},\ref{fig:branches}). However, Figure~\ref{fig:hys} shows that the diffusivity parameter of water vapor through the inhibition layer, $K_{zz}$, clearly also influences the surface climate. Moreover, at some values of $K_{zz}$ multistability is negligible, while at other values it is large. Why?

Figure \ref{fig:diff_comp} shows the vertical profiles of hot atmospheres that are all in the inhibited regime. We find that the diffusivity $K_{zz}$ has a strong but non-monotonic impact on surface climate. With a low $K_{zz}$ of $10^{-4} \unit{m^2/s}$, the inhibition layer is very shallow and occupies only one model layer (Figure \ref{fig:diff_comp}(a)). As $K_{zz}$ increases to $0.2\unit{m^2/s}$, the inhibition layer becomes significantly thicker and surface temperature rises about 40~K. However, if $K_{zz}$ increases even more, the surface temperature drops again. At a very high $K_{zz}$ of $20\unit{m^2/s}$, the superadiabatic temperature jump in the lower atmosphere essentially disappears, and the atmosphere closely resembles a moist adiabat.

The non-monotonic dependence of surface temperature on $K_{zz}$ arises from two competing effects. At low $K_{zz}$, the specific humidity is high at the surface and low above. The inhibition layer is shallow, the relative humidity above the inhibition layer is low, and the convective atmosphere immediately above the inhibition layer follows a dry adiabat (see Figure \ref{fig:diff_comp}(a); inside the layer of dry convection, specific humidity is vertically uniform). An increase in $K_{zz}$ then increases the specific humidity in the free atmosphere, which pushes a larger fraction of the atmosphere above the Guillot threshold. As a result, the thickness of the inhibition layer increases, which in turn also requires a hotter surface temperature.
However, at high $K_{zz}$, the diffusion of water vapor results in a vertical latent heat flux which starts to out-compete the upward radiative heat flux through the inhibition layer. To see why, first consider why the onset of convective inhibition results in a large temperature jump near the surface: once convection is suppressed, the surface cannot effectively shed heat via convection anymore. The only way left for it to get rid of its heat is via radiation, and the upward radiative heat flux is larger if the vertical temperature gradient is larger \citep{raymond_t_pierrehumbert_principles_2010}. However, as $K_{zz}$ increases, the sub-scale moisture diffusion also produces an upward latent heat flux, eliminating the need for a superadiabatic temperature gradient. As a consequence, at very high $K_{zz}$ (see Figure \ref{fig:diff_comp}(d)), the $T$-$p$ profile of the inhibition layer becomes almost indistinguishable from a moist adiabat.

The same reason explains why, in our numerical results, the occurrence and magnitude of multistability depend on $K_{zz}$ (see Figure \ref{fig:hys}). Simulations with higher diffusivity tend to have a thicker inhibition layer. A thicker inhibition layer allows an atmosphere to support more multiple equilibria, and moreover allows the equilibria to be more distinct from each other.
Therefore, hot hycean planets should exhibit the largest possible spread in surface climates when the atmosphere's diffusivity is neither too high (which lowers near-surface temperature gradients) nor too low (which results in shallower inhibition layers). Interestingly, the only 3D simulations so far which explicitly analyzed the diffusivity of the inhibition layer suggest that $K_{zz}$ is right in the intermediate range in which these effects are the largest \citep{leconte_3d_2024}.

\section{Discussion and Conclusion}\label{sec:dis}

\subsection{Summary and Comparison against Previous Work}

We investigate the climates of hycean planets located within the habitable zone, focusing on the effects of moist convection inhibition. Our findings reveal that hycean planets exhibit complex climate dynamics, including bistability, oscillations, and multistability. Bistability and oscillations arise because the OLR at which a warming planet develops an inhibition layer is different from the OLR at which the inhibition breaks down on a cooling planet.
Oscillations are favored for planets with thin atmospheres, $\mathcal{O}(0.1)\unit{bar}$, while bistability is favored for planets with thicker atmospheres, $\mathcal{O}(1)\unit{bar}$ or more. Multistability arises on hot planets whose surface temperatures significantly exceed the Guillot threshold, so the inhibition layer is thick. Our results suggest most, perhaps even all, hot hycean atmospheres in this regime might exhibit some form of multistability.

In the oscillating regime, the atmosphere goes through an inhibited phase in which the upper atmosphere cools down and a convective phase in which the upper atmosphere warms up. This is similar to the mechanism proposed in \cite{li_moist_2015}, in which the period of oscillations is determined by the radiative time scale of the atmosphere. The mechanism differs from the intermittent storms discussed by \cite{clement_storms_2024}, in which convective outbursts are mainly limited by the accumulation of methane in the lower atmosphere. The oscillation regime we find potentially mimics the episodic convection regime found in 3D simulations by \cite{seeley_resolved_2025}; however, in our case oscillations occur only at low surface pressures of $\mathcal{O}(0.1)$ bar whereas \citeauthor{seeley_resolved_2025} considered surface pressures of $1$ bar. Moreover, in our case the oscillations show clear periodicity, whereas the episodic convection reported in \citeauthor{seeley_resolved_2025} is at best quasi-periodic. More numerical work and analytical theory are needed to explain the gap between 1D and 3D simulations.

In the bistable regime, bistability occurs over a stellar flux range of $\mathcal{O}(1-10) \unit{W/m^2}$. This is quite significant compared to the width of the hycean habitable zone, which is $\mathcal{O}(10-100) \unit{W/m^2}$ \citep{pierrehumbert_hydrogen_2011,innes_runaway_2023}. Moreover, bistability occurs near the inner edge of the hycean habitable zone. This is relevant for ongoing exoplanet observations because observations of transiting exoplanets are biased towards shorter-period planets, which are more likely to be located near the inner edge of the habitable zone.

In the multistable regime the inhibition layer's thickness and surface climate are no longer uniquely constrained by the planet's TOA energy budget. Many different combinations of inhibition layer thickness and surface temperature can thus be in stable equilibrium with a given stellar flux. Some of our simulations even have multiple inhibition layers, which acts as if they are a single thick inhibition layer with thickness equal to their sum.
However, our 1D simulations also highlight additional considerations. The magnitude of multistability depends on the assumed sub-scale diffusion of water vapor through the inhibition layer; interestingly, multistability is largest right in the $K_{zz}$ regime suggested by 3D simulations \citep{leconte_3d_2024}. Perhaps not surprisingly, the number of stable equilibria in our 1D simulations further depends on the model's vertical resolution.

Extrapolating from our 1D results, one might expect even more complex climate dynamics on hycean planets in 3D models. For example, in our simulations multistability results in distinct stable equilibria but there is no guarantee that these states also have to be stable in 3D. Perturbations induced by atmospheric turbulence or other noise might induce the inhibition layer to jump between different stable states. These jumps might mimic noisy climate oscillations, such as the episodic convection reported by \cite{seeley_resolved_2025}; more work with 3D simulations is required to investigate this possibility.

\subsection{Implications for Hycean Planets}

Our results show that hycean climates are far more complex than is currently appreciated. This has consequences for the interpretation of sub-Neptune telescope observations, hycean planets, and the hycean habitable zone, which so far have been largely investigated using 1D models that ignore the effects of convective inhibition \citep[e.g.,][]{piette_temperature_2020,madhusudhan_interior_2020,madhusudhan_habitability_2021,nicholls_agni_2025}.

Previous work argued that the chemical characterization of sub-Neptunes can be used to confirm the existence of liquid water oceans \citep{yu_how_2021,hu_unveiling_2021,tsai_inferring_2021,hu_water-rich_2025}. 
The idea is that the observable abundances of species such as CO$_2$, CH$_4$, and NH$_3$ in the upper atmosphere depend on the atmosphere's chemical state, which is different for planets with deep gaseous atmospheres versus planets with shallow atmospheres and liquid $\wa$ oceans.
However, these investigations used 1D models which assumed that the atmosphere is in radiative-convective equilibrium, and hence well-mixed.
In reality, our work highlights that convective inhibition layers can pose strong barriers to mixing, so the atmospheres of hycean planets might not reflect their planets' interior bulk chemistry and composition. In our 1D model mixing through the inhibition layer is parameterized via a diffusivity $K_{zz}$, but in reality, mixing between the surface and upper atmosphere is highly uncertain and depends on both small-scale turbulence as well as large-scale atmospheric circulations. More work is thus needed with 3D models to explore the magnitude as well as possible spatial inhomogeneities in large-scale vertical mixing on sub-Neptunes.

Similarly, there are ongoing efforts to characterize the atmospheres of potential hycean planets such as K2-18b \citep{madhusudhan_habitability_2021}. Recent JWST observations of K2-18b even claimed to have detected DMS, a potential biosignature gas \citep{madhusudhan_carbon-bearing_2023}, although follow-up analyses cast strong doubt on this claim \citep{taylor_are_2025,stevenson_k2-18b_2025}.
So far it is still unclear whether K2-18b is actually a hycean planet. The planet's instellation is high, which means that it would require an unusually high albedo to still be habitable and sustain a liquid $\wa$ ocean \citep{pierrehumbert_runaway_2023,schmidt_comprehensive_2025,jordan_planetary_2025}. Should K2-18b indeed be habitable, it would likely be located near the inner edge of its habitable zone, and its atmosphere would likely be in the bistable or multistable regime. In these regimes, our results show that the planet's surface temperature could easily vary by at least 10-50~K depending on initial conditions and $K_{zz}$.
Additional work is thus needed to confirm whether previous claims about K2-18b are robust, such as extensive parameter and initial condition sweeps in 1D models that capture the effects of convective inhibition, combined with local and global 3D simulations \citep{leconte_3d_2024,barrier_new_2025}.

\subsection{Potential Shortcomings of Our Numerical Model}\label{sec:shortcoming}

\begin{itemize}
    \item We neglect reevaporation and allow all precipitation to reach the surface. This approximation probably makes it easier for the upper atmosphere to develop a subsaturated layer above the inhibition layer, making bistability more significant (see Figure~\ref{fig:AnaRegimes}).
    \item The surface climate presumably also depends on reevaporation. If large amounts of precipitation reevaporate as they pass through the inhibition layer, this will increase the amount of energy that has to be carried upward by radiation: in radiative equilibrium, the temperature gradient in the inhibition layer is roughly proportional to the net upward longwave flux, which equals the net downward shortwave flux minus the upward latent heat flux.
Reevaporation reduces the upward latent heat flux. To compensate, the upward radiative flux has to increase, which means surface temperature increases.
    \item Our model only considers moisture diffusion, but sensible heat diffusion could also affect the temperature structure of the inhibition layer. Including sensible heat diffusion should generally reduce the temperature lapse rate within the inhibition layer, but it would make our results even more dependent on the assumed value of $K_{zz}$.
    \item Our model assumes all surface fluxes are deposited in the lowest layer of the atmosphere. As a result, the inhibition layer in many of our simulations is only one model layer thick, presumably amplifying the small-scale noise seen in the inhibited states in Figure~\ref{fig:TS-cbd}. In reality, the thickness of the inhibition layer should not only depend on the magnitude of surface fluxes but also small-scale turbulence and large-scale atmospheric circulations.
    \item The period of oscillations, shown in Figure \ref{fig:AnaRegimes}, might be shorter than $\mathcal{O}(100)$ days. This is because, for oscillating simulations, we manually reduce surface fluxes by lowering the value of $w$ to avoid extremely short cycles. If the surface's effective heat capacity is large (e.g., if the ocean's mixed layer is deep), the period of oscillations would be mainly determined by heat transfer between the lower atmosphere and surface. The surface temperature remains high during the convective phase and surface fluxes rapidly heat up the bottom of the atmosphere. This can quickly stop convection and lead to short cycles. Conversely,  if the surface's effective heat capacity is small compared to the atmosphere's (e.g., if the ocean's mixed layer is shallow and the atmosphere is thick), it takes the atmosphere a relatively long time to heat up during the inhibited phase, so the period is long. However, a more detailed analysis requires quantitative knowledge of the thickness of the inhibition layer, which is beyond the scope of this work.
\end{itemize}

\bibliographystyle{aasjournal}
\bibliography{REFS_hyceanBis}

\end{document}